\documentclass[reprint,aps,prl,superscriptaddress]{revtex4-2}
\usepackage{bm,physics,amsmath,amssymb,
	graphicx,xcolor,embedfile,comment,mathrsfs,mathtools,xfrac,lipsum,hyperref}
\usepackage[utf8]{inputenc}
\usepackage[normalem]{ulem}

\usepackage[T1]{fontenc}
\usepackage{tikz}

\graphicspath{{figures/}}
\usepackage{float}

\makeindex

\begin{document}

\title{Interferometric determination of intrinsic nonlinear Kerr index in lead-halide perovskites}

\author{Dusan Lorenc}
\affiliation{Institute of Science and Technology Austria,
    Am Campus 1, 3400 Klosterneuburg, Austria}
\affiliation{International Laser Centre,
    Ilkovicova 3, 84104 Bratislava, Slovakia} 

\author{Ayan Zhumekenov}
\author{Osman M. Bakr}
\affiliation{KAUST (King Abdullah University of Science and Technology), Thuwal 23955, Saudi Arabia } 
      
\author{Zhanybek Alpichshev}
\email{alpishev@ist.ac.at}
\affiliation{Institute of Science and Technology Austria,
    Am Campus 1, 3400 Klosterneuburg, Austria} 


\begin{abstract}

Lead halide  perovskites have recently been reported to demonstrate an exceptionally high nonlinear (Kerr) refractive index n$_2$ of up to 10$^{-8}$ cm$^2$/W in CH$_3$NH$_3$PbBr$_3$. Other researchers however observe different, substantially more conservative numbers. In order to resolve this disagreement the nonlinear Kerr index of a bulk sample of lead halide perovskite was measured directly by means of an interferometer. This approach has many advantages as compared to the more standard z-scan technique. In particular this method allows studying the induced changes to the refractive index in a time-resolved manner, thus enabling to separate the different contributions to $n_2$. The extracted n$_2$ values for CsPbBr$_3$ and MAPbBr$_3$ at $\lambda \approx 1\mu m$ are $n_2=+2.1\times$ 10$^{-14}$ cm$^2$/W and $n_2=+6\times$ 10$^{-15}$ cm$^2$/W respectively hence substantially lower than what has been indicated in most of the previous reports implying the latter should be regarded with a great care. 

\end{abstract}

\maketitle


Lead halide perovskites (LHPs) remain at the forefront of a number of research areas including but not limited to photovoltaics, material- and condensed matter- physics and photonics, where LHPs were recently demonstrated to exhibit remarkable nonlinear optical properties (see e.g. \cite{Findik2021, Ferrando2018, Zhou2020}). In the specific case of lead-bromide perovskites  APbBr$_3$, a significant Kerr nonlinearity as high as n$_2 \sim 10^{-8}$\,cm$^2$/W \cite{Yi2017, serna2018nonlinear, Zhang2016, Kalanoor2016, Krishnakanth2018, Surez2019, Mirershadi2016} was reported, potentially making these compounds some of the most nonlinear bulk materials known to date. If confirmed, this can position LHPs as the future material of choice for $\chi^{(3)}$ nonlinear applications \cite{Zhou2020}. However, in addition to these reports of extreme Kerr indices there are also works that report much more conservative numbers for Kerr nonlinearity in that lead-bromide perovskites n$_2 \sim 10^{-14}$\,cm$^2$/W \cite{Kriso2020}.


Such orders-of-magnitude discrepancy across results obtained on nominally identically materials such as CH$_3$NH$_3$PbBr$_3$, implies that it is rather unlikely that the source of the variance in magnitude of nonlinearity is purely due to actual intrinsic material properties of respective LHP samples used in the studies. 

Besides material properties, the only remaining source of disagreement among Kerr index reports would be the variance in experimental details in different works, such as the character of light sources used and the geometries of the samples. It is important in this regard to mention that to the best of our knowledge majority of the n$_2$ values in LHPs reported so far were obtained via the so-called z-scan technique \cite{Zhou2020}. This method was developed in early 1990s and has since established itself as a default tool of choice for measuring Kerr index in bulk materials \cite{SheikBahae1990}. The technique relies on the fact that space-dependent local intensity of the probe beam via Kerr effect gives rise to a space-dependent local refractive index distribution inside the sample. Such profile manifests itself as a distortion of the wave-front of the probing beam. By modifying the intensity distribution of the probe and monitoring the effects on the probe beam, one can in principle deduce the value of Kerr index. Specifically, in z-scan the local intensity profile is modified by moving the sample along the beam direction through the focal region of the probe. The biggest merit of this method that has warranted its popularity, is its simplicity both in terms of experimental infrastructure and theoretical analysis. However, the description of the z-scan above also highlights the typical issues that one can face when employing this method. Firstly, since at different stages of the measurement different parts of the sample are illuminated, the local geometry of the sample and its optical uniformity along with the properties of the pump beam become crucial \cite{Chapple1997, Nalda2002}. Secondly, and most importantly, z-scan is essentially a static method that measures the integrated effect of the probing beam on the refractive index of the sample. In particular, in addition to the narrowly defined quasi-instantaneous ``intrinsic'' electronic Kerr effect, z-scan also picks up refractive index modifications due to ``extrinsic'' phenomena such as electrosctriction, thermo-optic and photo-refractive effects. 

\begin{figure}
  \includegraphics[width=\linewidth]{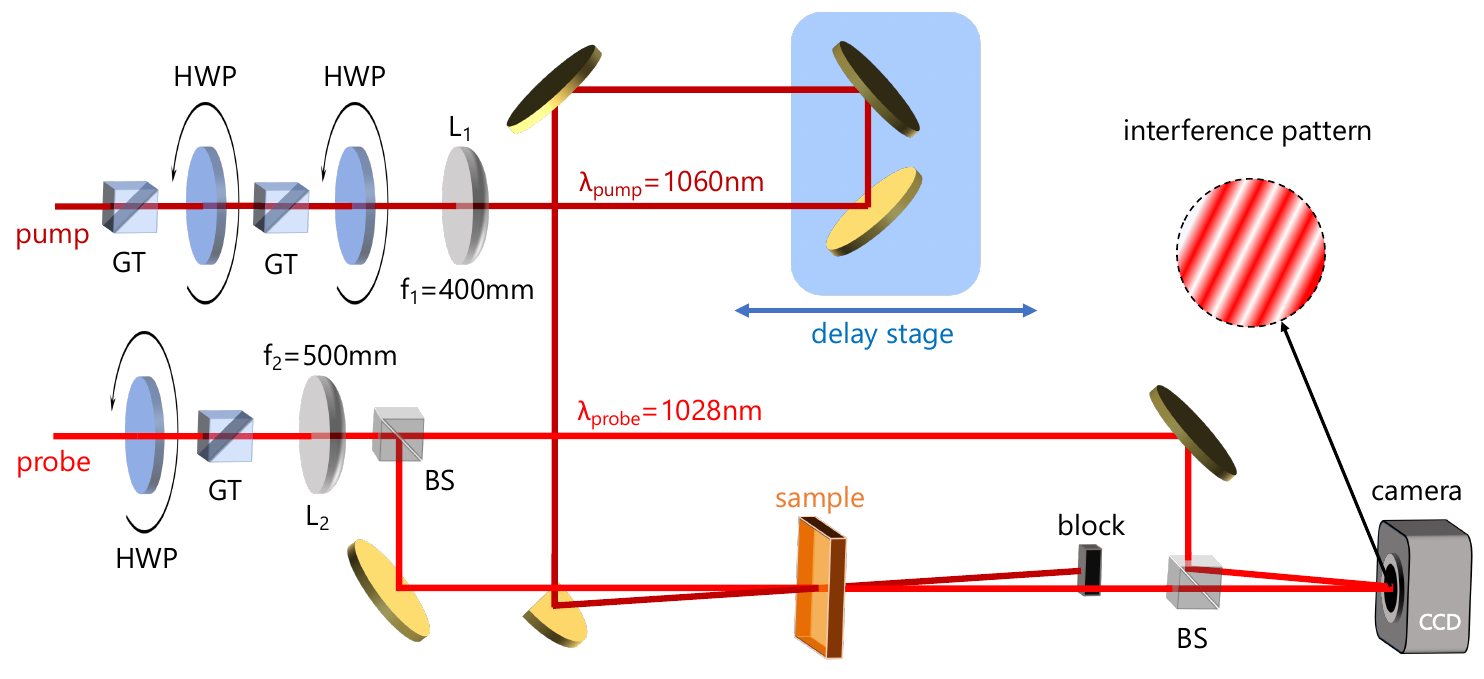}
  \caption{Schematic layout of the time-resolved Mach-Zehnder interferometer (MZI). Pump and probe beams are near-degenerate ($\lambda=$ 1060nm and 1028nm respectively). Transient pump-induced changes in refractive index are detected as shift in the interference pattern recorded by a CCD camera.}
  \label{fig:interferometer_w_sample1}
\end{figure} 

As it turns out, many of these issues are featured prominently in lead-halide perovskites. In particular, they have been demonstrated to exhibit strong photo-refractive and electrostrictive effects \cite{Chen2018, Tahara2018}. Thermo-optic corrections to the refractive index were also found to be featured prominently in LHP samples \cite{Handa2019}, which becomes of particular concern given the exceptionally poor heat conductivity of these materials \cite{Haeger2020}. The main problem here is that extrinsic contributions to the Kerr response are orders of magnitude stronger than the electronic one \cite{Boyd2008, Chang1981}, making it challenging to extract the intrinsic nonlinear refractive index of the medium from studying the self-interaction of a light beam under general circumstances. To account for this, it is conventional to perform experiments using ultrafast laser pulses and make use of the quasi-instantaneous nature of the electronic response. The works that implement z-scan measurements typically check for fluence/repetition rate dependence of their z-scan signals \cite{Gnoli2005} in order to decouple long-lived extrinsic effects from the instantaneous intrinsic ones, however this only works when the unwanted phenomena occur at timescales comparable to the typical pulse repetition period. When the extrinsic effects persist for significantly longer than the longest interval between pulses, it is not always possible to isolate the instantaneous response (for instance the photorefractive effect may have response time on the order of tens of seconds).

\begin{figure}[t]
  \includegraphics[width=\linewidth]{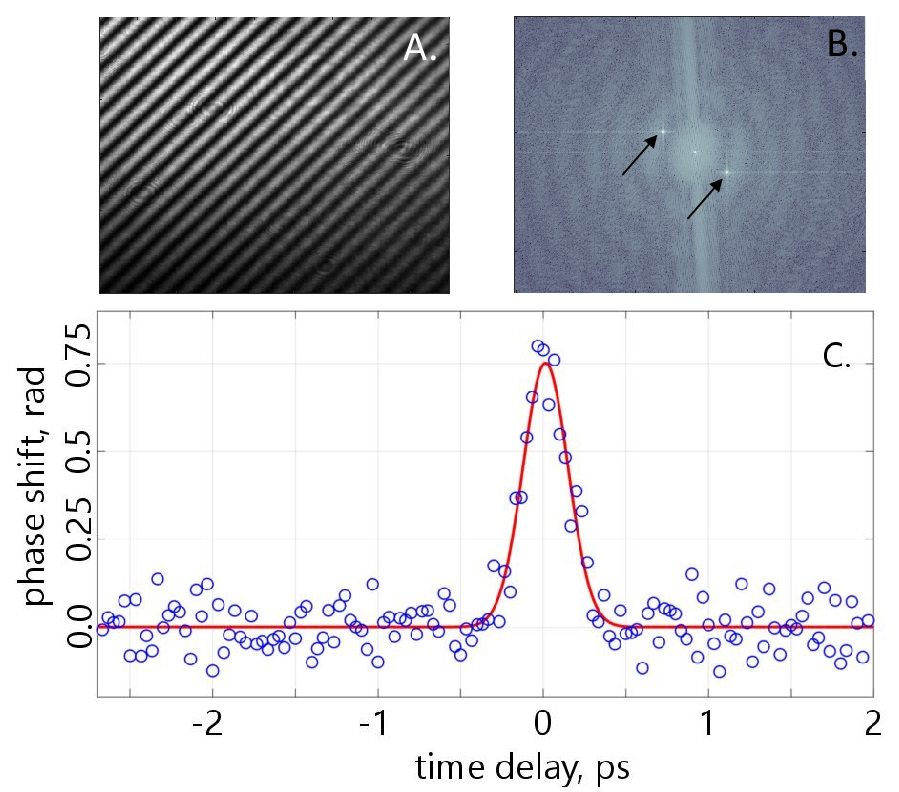}
  \caption{A) sample interferogram as recorded by CCD camera; B) 2D Fourier intensity of the interferogram in A). The real-space shift of the interference pattern is inferred from the change of the phase of complex amplitude at the peak. Time dependence of the pump-induced phase shift between the two probe beams in MZI; C) Pump-induced transient phase shift of the interference pattern in single crystal CsPbBr$_3$ taken with a pump pulse with peak intensity $I_0=3\times10^{9}$W/cm$^2$. Solid red line is a fit to a gaussian curve to extract the peak value of the phase shift $\Delta \phi$ at this pump intensity. }
  \label{fig:CsPbBr3}
\end{figure}

Here we circumvent the complications associated with z-scan measurements of the intrinsic (instantaneous) nonlinear Kerr index in LHP by means of measuring n$_2$ in a time-resolved fashion. To this end we measure the nonlinear corrections to the refractive index directly by means of interferometry. As will be demonstrated below, such approach has a number of advantages as compared to the classic z-scan method: 1) the setup remains static throughout measurement, i.e. sample is not being moved, therefore the method is insensitive to the irregularities of the geometry and the inner inhomogeneities of the sample; 2) the average incident power on the sample remains the same throughout the experiment, the heating due to radiation thus only affecting the effective sample temperature; 3) studying transient changes of refractive index in real time allows separating different contributions to the nonlinear response; 4) pump and probe are separate beams whose polarizations can be controlled independently; 5) not being dependent on sample homogeneity the proposed approach allows measuring on bulk single crystal samples. This has the advantage of minimizing the possible size effects in nano-crystals and to focus on the intrinsic properties of the material.

\section{Experiment}

In optical Kerr effect, the (phase) refractive of the medium is modified in the presence of radiation with intensity $I$ according to $n(I) = n_0 + n_2 I$ with $n_0$ and $n_2$ are referred to as linear nonlinear (Kerr) refractive indices respectively. While it is often the case that $I$ is the intensity of the very beam that also ``experiences'' the refractive index $n$ (self-phase modulation), it does not always have to be so. In this work, to study the Kerr effect in a time-resolved manner we use two separate laser beams. The first, more intense one (pump) is used to provide the intensity $I$ to modify the refractive index of the medium while the other one (probe) is used to probe the resulting refractive index. Since Kerr-induced changes $\Delta n=n_2 I$ are usually weak for generic materials, we choose to detect them here by resorting to interferometry. 

The experimental layout is schematically shown in Fig.~\ref{fig:interferometer_w_sample1}. It consists of a Mach-Zehnder interferometer where the refractive index $n$ of a sample put into one of the arms is modified by the pump beam. This change in $n$ results in the change of the optical length of the corresponding arm of MZI, which manifests itself as a shift in the interferogram recorded by a CCD camera located at the exit of MZI. The delay between the pump and probe pulses can be controlled by a fast motorized delay stage (Newport {\it LTLMS800}). To extract the time dependence of transient changes to $n$, the stage is moved incrementally and an interferogram is recorded for each position of the stage. In order to avoid the detrimental effect of mechanical vibrations, for every delay stage position we record two consecutive intefrograms: one at the actual desired position and the other at a fixed reference position predating pump arrival. The actual phase shift is then recorded as the difference in position of the interferogram relative to the one taken at the reference position. Together with the high speed of the delay stage (500mm/s) this method helps minimize the effect of the low frequency noise coming from mechanical deformations of MZI.  Each interferogram is then processed using a 2D fast Fourier transform and subsequently evaluated in order to extract the corresponding amplitude and phase (see Fig.2B,C).

\begin{figure}
  \includegraphics[width=\linewidth]{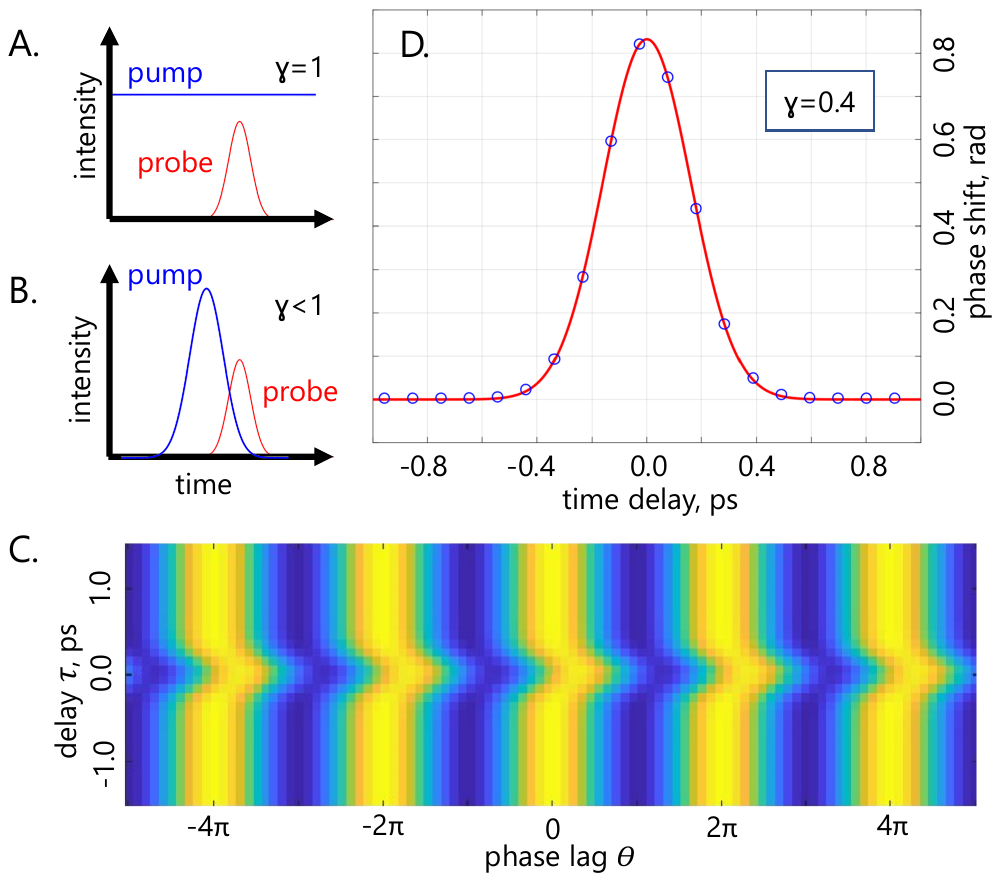}
  \caption{A) Default Kerr-effect configuration: the pump intensity is constant, and $\gamma=1$ (see text); B) Experimental configuration with pulse-shaped pump and probe profiles. Different parts of the probe pulse experience different instantaneous pump intensity $I(t)\leq I_0$, therefore $\gamma < 1$; C) CGNLSE-simulated interference pattern as a function of pump-probe delay $\tau$ calculated for CsPbBr$_3$ at $I_0=3\times10^{9}$W/cm$^2$ and $n_2= +2.1\times$ 10$^{-14}$ cm$^2$/W; D) Pump-induced phase shift of interference pattern extracted from C) versus delay $\tau$ (blue dots) and a gaussian fit to it (red line).}
\label{fig:simulation}
\end{figure}

An amplified femtosecond laser system (Light Conversion {\it PHAROS}) coupled to an optical parametric amplifier (OPA, Light Conversion {\it ORPHEUS}) is used as the principal laser source. The laser produces a train of pulses centered at 1028nm with a repetition rate of 3kHz, pulse duration of 300fs and a pulse energy of 2mJ. A small part of the beam (5\%) has been coupled into the Mach-Zehnder interferometer (MZI) through a variable attenuator consisting of a half-wave plate (HWP) and a Glan-Taylor polarizer; while the remaining part pumped the OPA tunable between $(0.66 - 16)\mu$m. In this work, the fundamental wavelength of 1028nm was used as a probe while the 1060nm idler output has been used as the pump radiation. The pump wavelength was chosen such that is nearly degenerate with probe, but is still different enough to be effectively separable from the latter. The delay between pump and probe is controlled with a linear stage. Both pump and probe passed through a half-wave plates (HWP) and polarizers to attenuate their respective powers and to clean the polarizations. The pump beam then passes another free standing HWPs in order to set its polarization state. The pump and probe beams were loosely focused and overlapped in the sample in a slightly non-collinear arrangement. Care was taken to make sure that the probe diameter inside of the sample is considerably smaller than that of the pump. This is to ensure that pump intensity is a well defined quantity. The probe light leaving the output port of MZI passed through a $1030\pm5$\,nm bandpass filter and the resulting interferograms were recorded by a CCD camera (Point Grey Research Inc. {\it CM3-U3-50S5M}). The pump intensity is tunable in a broad range while the probe intensity was kept fixed at $I_{probe}=4\times10^6$\,W/cm$^2$. High quality bulk single crystal samples of CH$_3$NH$_3$PbBr$_3$ were grown by inverse temperature crystallization method as described elsewhere \cite{Saidaminov2015, Saidaminov2015b} and similarly high quality bulk single crystal samples of CsPbBr$_3$ were grown by antisolvent vapor-assisted crystallization method as described in \cite{Rakita2016}. All experiments were performed at room temperature {\it i.e.} in the orthorhombic phase of CsPbBr$_3$ and cubic phase of MAPbBr$_3$.

\begin{figure*}
  \includegraphics[width=0.9\linewidth]{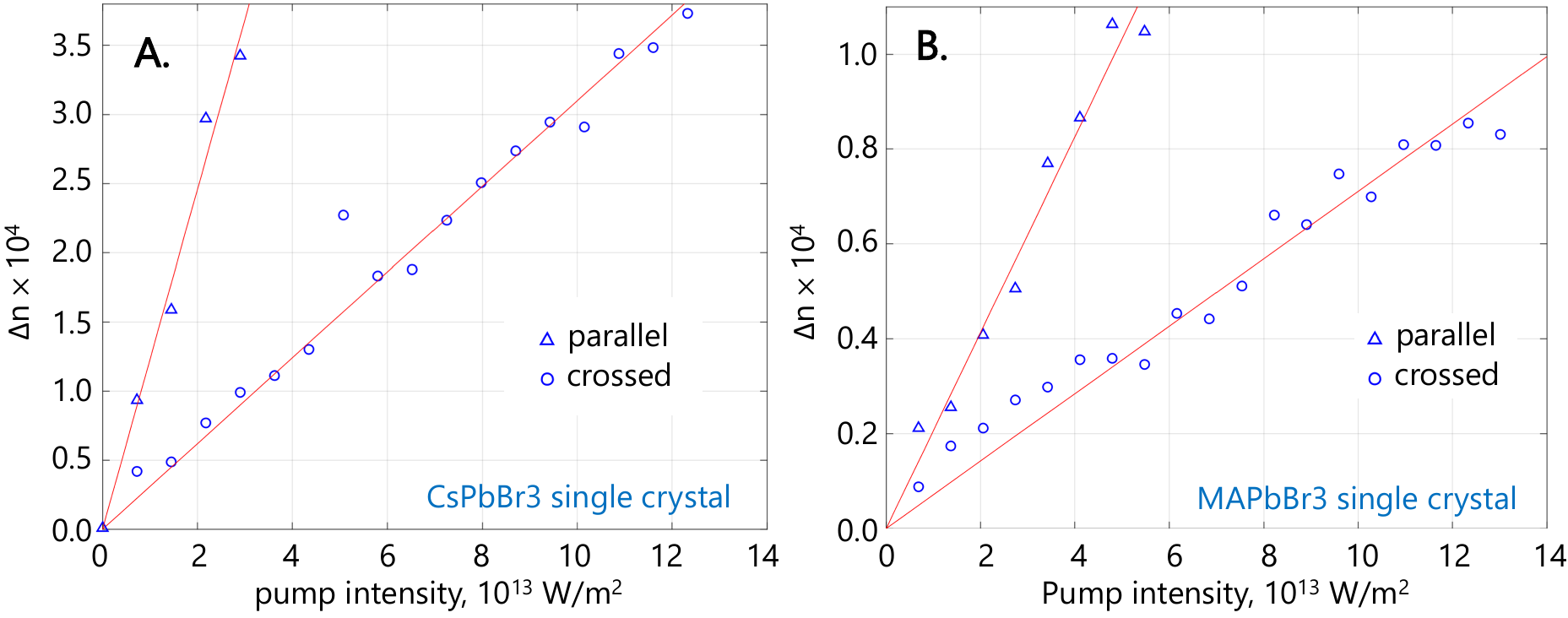}
  \caption{Experimental Kerr-induced peak refractive index change $\Delta n$ as a function of peak pump intensity for A) CsPbBr$_3$; and B) MAPbBr$_3$ for parallel and crossed polarization configurations. Nonlinear (Kerr) refractive index $n_2$ is found from fitting (red lines) as a coefficient of linear relation $\Delta n= n_2 I$.}
  \label{fig:ref_index}
\end{figure*}

\section{Results and Analysis}
Fig.~\ref{fig:CsPbBr3}c shows a sample trace of the transient phase shift as a function of pump-probe delay for the CsPbBr$_3$ sample taken at peak pump intensity $I_0=3\times10^{9}$W/cm$^2$. The first thing to notice here is that the temporal width of the signal is comparable to pulse-width of the pump pulse ($\tau \approx 270$\,fs) meaning that the phase response can be considered instantaneous, and therefore can be fully ascribed to the intrinsic electronic hyperpolarizability of the medium. Having established the nature of the response this way, we can use the data to determine the numeric value of the intrinsic Kerr nonlinear index $n_2$. To this end however one needs to establish a relation between the measured peak phase shift $\Delta \phi_0$ and the Kerr-induced change in refractive index $\Delta n = n_2 I_0$. Based on scaling considerations, it is clear that one can write:

\begin{equation}
\Delta \phi_0 = \gamma \times n_2 I_0 \, \,  2 \pi  L/\lambda
\label{eq:gamma} 
\end{equation}

where $L$ and $\lambda$ are the thickness of the sample and probe wavelength respectively; $I_0$ is the peak intensity with an unknown coefficient $\gamma \lesssim 1$ whose exact value depends on the particular details of pump and probe pulse temporal profiles ($\gamma =1$ in the limit $I_{\mathrm{pump}}(t)=I_0$; see Figs.\ref{fig:simulation}A and B). To estimate the value of $\gamma$ we simulate the Kerr-interaction between the pulses by means of integrating numerically a system of coupled generalized nonlinear Schroedinger equations (CGNLSE) for frequency-nondegenerate fields \cite{Agrawal2013}. 

\begin{align}
\begin{split}
\frac{\partial A_{1}}{\partial z}+\frac{1}{v_{g1}}\frac{\partial A_{1}}{\partial z}+i\frac{\beta_{21} }{2}\frac{\partial^2 A_{1}}{\partial z^2} &= i\Gamma_{1}(\left | A_{1} \right |^{2}+2\left | A_{2} \right |^{2})A_{1}\\
\frac{\partial A_{2}}{\partial z}+\frac{1}{v_{g2}}\frac{\partial A_{2}}{\partial z}+i\frac{\beta_{22} }{2}\frac{\partial^2 A_{2}}{\partial z^2} &= i\Gamma_{2}(\left | A_{2} \right |^{2}+2\left | A_{1} \right |^{2})A_{2} 
\end{split}
\label{eq:CGNLSE}
\end{align}

\noindent Here $A_{j}$ stands for the field amplitude of $j^{th}$ beam ($j=1,2$ for pump pump and probe beams respectively); $v_{gj}$ for the corresponding group velocities and $\beta_{2j}$ are the GVD coefficients. These parameters can be extracted from refractive index measurements in these compounds \cite{Volosniev2023, Wei2023, Ermolaev2023}. Finally, $\Gamma_{j}$ are the third-order nonlinear parameters that determine $n_2 = \Gamma_{j} \, \lambda_j S^j_{\mathrm{eff}}/2\pi $,  here $\lambda_j$ denoting the wavelength of the corresponding beam and $S^j_{\mathrm{eff}}$ being its effective diameter as defined in \citep{Agrawal2013}.

In order to relate the resultant pump-modified probe field amplitude $A_p(\tau, t)$ at the exit of the sample calculated for pump-probe delay $\tau$ to the experiment, we interfere it with a reference field $A_{0}(t)$ (calculated by integrating the same system of equations eq.\ref{eq:CGNLSE} with pump intensity put equal to zero) and calculate the resulting intensity. 

\begin{equation*}
F(\theta, \tau)= \! \! \! \int\displaylimits_{-\infty }^{+\infty } \! \!  \left[A_{p}(\tau, t)A_{0}^{*}(t)e^{i\theta}  +A_{p}^{*}(\tau, t)A_{0}(t) e^{ -i\theta} \,\right] dt 
\label{eq:interference}
\end{equation*}

\noindent Here we introduce phase lag $\theta$ between the two fields. This quantity $F(\theta, \tau)$ above has the meaning of being proportional to the net average power resulting from the interference between the probe and reference beams as a function of phase lag $\theta$ between the interferometer arms and pump-probe delay $\tau$. By calculating $F(\theta, \tau)$ one can simulate the Kerr-induced shift of the interference pattern, as shown in Fig.\ref{fig:simulation}C. Tracing the position of maximum of $F(\theta, \tau)$ as a function of $\tau$ for given $n_2$ and $I_0$ and comparing the peak phase shift $\Delta \phi$ with eq.\ref{eq:gamma} we find that for the parameter values of our experiment  $\gamma_{\mathrm{Cs}} \approx 0.4$ for CsPbBr$_3$ and $\gamma_{\mathrm{MA}} \approx 0.7$ for MAPbBr$_3$.   

By using this values of $\gamma$, we can now convert phase shifts $\Delta \phi$ to refractive index changes $\Delta n$ and plot the peak values of the latter as a function of peak pump intensity $I_0$ for both CsPbBr$_3$ and MAPbBr$_3$ as shown in Fig.\ref{fig:ref_index}. To characterize the nonlinear Kerr effect in our LHP samples we probe it in two configurations where pump and probe polarizations were kept parallel (``$xx$'') and crossed (``$xy$'') with respect to each other; in both cases the polarization planes are aligned with the crystal axes. Here we can see that refractive index follows linear dependence $n(I)=n_0+n_2 I$. From these fits we obtain for CsPbBr$_3$ $n_2^{xx}=(2.1\pm0.2)\times10^{-14}$\,cm$^2$/W (pump and probe polarizations parallel) and $n_2^{xy}=(5.5\pm0.2)\times10^{-15}$\,cm$^2$/W (crossed polarizations). For MAPbBr$_3$ we obtain $n_2^{xx}=(6.0\pm0.2)\times10^{-15}$\,cm$^2$/W and $n_2^{xy}=(3.3\pm 0.1)\times10^{-15}$\,cm$^2$/W 
These numbers are rather modest in comparison to some of the previous reports of $n_2$ in LHPs \cite{Zhou2020}, on the other hand, by the order of the magnitude, they nicely fall into the range expected for generic semiconductors with comparable band gap, such as ZnSe \cite{SheikBahae1990a,SheikBahae1991,Lorenc2023}. 

\section{Conclusion}

In summary we measured the intrinsic nonlinear (Kerr) refractive index $n_2$ of a bulk single crystal CsPbBr$_3$ and MAPbBr$_3$. Given the orders-of-magnitude variance among reported Kerr index values in these compounds, which to the best of our knowledge were obtained by means of z-scan technique, we chose to measure $n_2$ by means of an alternative method. Namely, we have employed a direct interferometric pump-probe technique. Being a time-resolved alternative to the classic z-scan technique, it has a number of advantages as compared to the latter, the most important one being that it can measure the instantaneous response of the medium while in z-scan one works with a time-integrated signal. This distinction becomes important whenever extrinsic, non-electronic channels of Kerr nonlinearity, such as photo-refractivity or heat-related phenomena, are characterized by long time-scales as compared to the repetition rate of the probing source.

Another advantage of the method adopted in this work, are the less stringent requirements imposed on the sample geometry as compared to z-scan, which allows us to measure $n_2$ in bulk single-crystal samples, thus minimizing the possible surface- and size-effects in the grains of polycrystalline samples. By analysing the pump-induced phase shifts of the probe beam we obtain $n_2=+2.1\times 10^{-14}$ cm$^2$/W for CsPbBr$_3$ and $n_2=+0.6\times 10^{-14}$ cm$^2$/W for MAPbBr$_3$. These numbers are significantly lower than the highest reported $n_2$ values in LHPs, on the other hand they are very reasonable for generic semiconductors with comparable electronic band gaps \cite{SheikBahae1990a, SheikBahae1991}.

\medskip

\medskip
\textbf{Acknowledgements} \par 
We gratefully acknowledge assistance of Prof. John Dudley.

\medskip

%
\bibliographystyle{MSP}
\bibliography{n2_perovskites}



\end{document}